\documentclass[a4paper,12pt]{article}
\usepackage{amssymb}

\usepackage{hyperref}
\usepackage{subfigure}
\usepackage{ae}
\usepackage{mathrsfs}
\usepackage{graphicx}
\usepackage{bbm}
\usepackage{latexsym}
\usepackage{dsfont}
\usepackage{amsmath}
\usepackage[title,titletoc]{appendix}
\usepackage[super]{nth}
\usepackage{cancel}
\usepackage{booktabs}
\usepackage[compat=1.0.0]{tikz-feynman}
\usepackage{multicol}
\usepackage{bbold}
\usepackage{amsfonts}
\usepackage{cite}

\newcommand{\mathsym}[1]{{}}

\newcommand{\qed}{\nobreak \ifvmode \relax \else \ifdim\lastskip<1.5em \hskip-\lastskip \hskip1.5em plus0em minus0.5em \fi \nobreak \vrule height0.75em width0.5em depth0.25em\fi}
\newcommand{\diag}{\mbox{diag}}

\def\app#1#2{  \mathrel{    \setbox0=\hbox{$#1\sim$}    \setbox2=\hbox{      \rlap{\hbox{$#1\propto$}}      \lower1.1\ht0\box0    }    \raise0.25\ht2\box2  }}

\begin{document}

	\begin{titlepage} 
		\begin{center} \hfill \\
		\hfill \\
			\textbf{\Large
Does Quark Mixing play a role in the Lepton Sector?}

			 \vskip 1cm \vskip 1cm Francisco Albergaria \footnote{francisco.albergaria@tecnico.ulisboa.pt}, G.C. Branco\footnote{gbranco@tecnico.ulisboa.pt},
José Filipe Bastos \footnote{jose.bastos@tecnico.ulisboa.pt} and
J.I. Silva-Marcos\footnote{juca@cftp.tecnico.ulisboa.pt}

\vskip 0.07in Centro de
F{\'\i}sica Te\'orica de Part{\'\i}culas, CFTP, \\ Departamento de
F\'{\i}sica,\\ {\it Instituto Superior T\'ecnico, Universidade de Lisboa, }
\\ {\it Avenida Rovisco Pais nr. 1, 1049-001 Lisboa, Portugal} \end{center}

		\begin{abstract} 
We suggest a simple relation between the quark and the lepton mixing within the framework of type-I seesaw
mechanism. We show that within our ansatz the empirical King-Mohapatra-Smirnov relation, which suggests a connection between the CKM and PMNS mixing where $|V^\text{PMNS}_{13}|\approx \frac{1}{\sqrt{2}}\sin(\theta_C)  $, can be derived.
This is possible within a restricted region of the Dirac and Majorana mass parameters.
\end{abstract}
	\end{titlepage}
	
\section{Introduction}
	
In the Standard Model (SM) gauge invariance does not constrain the flavour structure of Yukawa couplings and as a result quark masses and mixing are not predicted within the SM. The Yukawa couplings have to be adjusted to reproduce the observed quark masses and mixing. Furthermore, in the SM there is no relation between the quark and lepton sectors. Yet, the idea of having a relation between the quark and lepton sectors is very appealing and various attempts at establishing such a relation have been suggested in the literature, often referred to as quark-lepton symmetry\cite{Marshak:1979fm, Foot:1990dw, Achiman:1978vg, Clarke:2011aa, Foot:1991fk, Smirnov:1995jq}.

In the SM neutrinos are strictly massless, so the discovery that at least two of the neutrinos have non-vanishing masses rules out the SM. However, it is easy to have an extension of the SM where neutrinos acquire non-vanishing masses. The simplest extension consists of just introducing right-handed neutrinos. The most general gauge invariant Lagrangian also includes a right-handed neutrino Majorana mass term involving a matrix $M_R$. Since this mass term is gauge invariant, it can be much larger than the electroweak scale $v$, which is associated to the Dirac mass terms. This immediately leads to the seesaw mechanism, with the light neutrinos having a mass of order $m_D^2/m_R$, where $m_D$ and $m_R$ represent the Dirac and Majorana mass term orders respectively.

In this paper and in the framework of seesaw type one, we illustrate how a relationship between the quark and lepton sectors can be implemented. In the seesaw framework, without loss of generality, one may choose a weak basis where the charged lepton mass matrix and the right handed Majorana neutrino mass matrix are both diagonal and real. In this basis, all the mixing arises from the Dirac neutrino mass matrix $M_D$ which is arbitrary. Our Ansatz consists of assuming a lepton quark symmetry leading to the suggestion  that $M_D$ is related to $V_{\text{CKM}}$. We show that in our framework one obtains the King-Mohapatra-Smirnov (KMS) relation \cite{Smirnov2004,Antusch2005,King2012,Farzan:2006vj}, connecting the elements of the CKM and PMNS matrices.

The paper is organised as follows: in the next section, we present our framework leading to the King-Mohapatra-Smirnov relation. In section three, we give a concrete numerical example of our ansatz which exhibits the proposed features. We also present a numerical analysis which explores the parameter-region near to the points of our ansatz. Finally, in section four, we state our conclusions.
	
\section{Framework}


Consider the fermion mass matrices in the SM. With respect to the quark sector, we recall that one may choose a weak basis in which the mass matrices of the up- and down-type quarks have the form
\begin{equation}
    M_u = D_u, \qquad M_d = V D_d  V^\dagger,
\end{equation}
where $V$ stands for the CKM quark mixing matrix $V_{\text{CKM}}$, and $D_u, D_d $ are diagonal matrices representing the quark masses, $ D_u= \diag (m_u, m_c, m_t)$,  $D_d = \diag (m_d, m_s, m_b)$. This weak basis contains all physical relevant parameters.

Now, consider the lepton sector in the context of the type-I See-Saw framework with $3$ right-handed neutrinos. The mass terms of this sector are given by
\begin{equation}
    \mathcal{L}_m=-\left[\overline{\nu}_L M_D \nu_R + \frac{1}{2}\nu^T_R C M_R \nu_R + \overline{\ell}_LM_\ell \ell_R \right]+h.c.,
    \label{lagrangian}
\end{equation}
where $\nu_{L,R}$ and $\ell_{L,R}$ are the neutrino and charged lepton flavour eigenstates, respectively. For the charged leptons we have a mass matrix $M_\ell$, and for the neutrinos, we have a Dirac mass matrix term $M_D$ along with the Majorana mass matrix $M_R$. All of these are $3\times 3$ matrices.

Similar to the previously referred quark sector, in the lepton sector, one may choose a weak basis in which the mass matrix of the charged leptons is diagonal $M_\ell = D_\ell = \diag (m_e, m_\mu, m_\tau)$. One may also set $M_R $ the heavy right-handed Majorana neutrino mass matrix to be diagonal $M_R=D_R = \diag(M_1, M_2, M_3)$ (with, $M_1$, $M_2$ and $M_3$ being approximately the masses of the three heavy right-handed Majorana neutrinos), while the Dirac neutrino mass matrix $ M_D$ has the form:
\begin{equation} \label{eq:leptonmass}
  M_D = V' D_D W^\dagger, \qquad  M_\ell = D_\ell, \qquad  M_R = D_R,
\end{equation}
where $D_D = \diag(m_{D_1}, m_{D_2}, m_{D_3})$ with $m_{D_1}$, $m_{D_2}$ and $m_{D_3}$ the parameters corresponding to masses of the three Dirac neutrinos. The $3\times 3$ matrices $V'$ and $W$ are, for now, just general unitary matrices {\footnote{Although, in the end one has to obtain the correct light-neutrino masses and PMNS mixing}}.

\subsection*{Ansatz inspired by Quark-Lepton symmetry}

At this point, we suggest an analogy between the lepton sector and the quark sector. This analogy/relationship might originate from some (as yet unknown) quark-lepton symmetry or mechanism such that the matrices $V', W$ are close to the quark CKM mixing matrix $V=V_{\text{CKM}}$.
A priori, this proposal is a very confining, and it is not clear that it might lead, in some way, to any effective physical content.

Even more restrictive, we may focus on the limit situation putting forward the \textit{ansatz} possibility where we strictly have that $V^\prime = W = V(\equiv V_{\text{CKM}})$ in Eq. \eqref{eq:leptonmass} and try to work out if this is compatible with the known lepton masses and mixing. Later, in a more extensive numerical analysis we will explore a larger region of parameters where $V', W$ are merely close to the quark CKM mixing matrix.

In this exact case where $V^\prime = W = V(\equiv V_{\text{CKM}})$, we have
\begin{equation} \label{eq:leptonmass2}
    M_\ell = D_\ell, \qquad M_D = V D_D V^\dagger, \qquad M_R = D_R.
\end{equation}
Then, if the heavy right-handed Majorana masses $M_i$ ($i=1,2,3$) are sufficiently large, the effective neutrino mass matrix reduces to
\begin{equation}
    M_{\text{eff}} = - M_D M_R^{-1} M_D^T.
    \label{eq:eff}
\end{equation}
which with Eq. \eqref{eq:leptonmass2} can be written as
\begin{equation}
\begin{split}
    M_{\text{eff}} =& - V\ ( D_D V^\dagger\  D_R^{-1}\ V^{*} D_D)\ V^T \\
    =&-V\ M_0\ V^T.\\
   \end{split}
     \label{eq:eff2}
\end{equation}
where we have defined the sub mass matrix $M_0 \equiv D_D V^\dagger\  D_R^{-1} \ V^* D_D$. 

We thus find, that the effective neutrino mass matrix is diagonalized by
\begin{equation}
    U = V^* T^*,
\label{diag}
\end{equation}
where $T^*$ stands for the unitary matrix which diagonalizes the sub mass matrix $M_0 = D_D V^\dagger\  D_R^{-1} \ V^* D_D$. 

So far, what we have done is just to simply write in Eq. (\ref{diag}) the diagonalization of the effective neutrino mass matrix $ M_{\text{eff}}$, and which, for our \textit{ansatz}, has a contribution from the CKM matrix $V=V_\text{CKM}$. Next, we we shall take this a step further towards the intriguing phenomenological hypothesis connecting the quark-CKM and lepton-PMNS mixing: the so-called King-Mohapatra-Smirnov (KMS) relation.

\subsection*{The King-Mohapatra-Smirnov relation}

Proceeding with the sub mass matrix $M_0$, if this matrix is such that its diagonalization with $T$ is near to the tri-bimaximal mixing matrix, $T_{\text{TBM}}$, \textit{i.e.} if $T \sim T_{\text{TBM}}$, then we will obtain in approximation
\begin{equation}
  V^{\text{PMNS}} = V_\text{CKM}^* T_{\text{TBM}}^*,
\end{equation}
thus yielding the so-called King-Mohapatra-Smirnov relation which establishes the well-known match between $V^{\text{PMNS}}_{13}$ and the exact TBM result for atmospheric neutrino angle in connection with the quark Cabibbo angle, i.e.  
\begin{equation}
\begin{split}
   |V^{\text{PMNS}}_{13}| =& \left|\sum_{i=1}^3 (V_{\text{CKM}})_{1 i}  (T_{\text{TBM}})_{i 3} \right|
   \\
   = &\left| (V_{\text{CKM}})_{1 2}  (T_{\text{TBM}})_{2 3} + (V_{\text{CKM}})_{1 3}  (T_{\text{TBM}})_{3 3} \right|
   \\
   \approx &  \left| (V_{\text{CKM}})_{1 2}  (T_{\text{TBM}})_{2 3} \right|
   \\
   \approx &|\sin \theta_C  \frac{1}{\sqrt{2}}|\approx 0.22\  \frac{1}{\sqrt{2}} \approx 0.156,
\end{split}
\end{equation}
where we used $|(V_{\text{CKM}})_{12}| \gg |(V_{\text{CKM}})_{13}|$ and $|(T_\text{TBM})_{13}| = 0$, $|(T_\text{TBM})_{23}| =$ $ |(T_\text{TBM})_{33}|=\frac{1}{\sqrt{2}}$.

Obviously for this to be true, the matrix $M_0$ which depends on $V$, on the Dirac neutrino masses $m_{D_i}$, and on the Majorana neutrino masses $M_i$, must not only yield the correct light-neutrino masses $m_{\nu_i}$ but also its diagonalization matrix $T$ must be near to the tri-bimaximal mixing matrix, $T_{\text{TBM}}$.

A priori, one might think that this is not at all viable with the limited set of parameters at stake. However, searching through parameter-space, we have intriguingly enough found a region which makes this particular quark-lepton integration possible. 
This achievement also illustrates how small mixing, expressed in the CKM matrix $V$, may result in the lepton sector, through the see-saw mechanism, in large mixing given by the PMNS matrix.

Moreover, assuming that the Dirac neutrino masses are of the order of the EW scale or somewhat smaller, our framework makes specific predictions for the heavy Majorana neutrino masses, which must then be of the order of $10^{10}\ \text{GeV}$ or larger. 

\section{Numerical Analysis}

\subsection*{Example with Normal Ordering}

Next, we present a numerical example, showing that indeed it is possible for $M_0$ have the correct light-neutrino mass-eigenvalues $m_{\nu_i}$ (and thus also $M_{eff}$) and at the same time to be diagonalized by a matrix $T$ near to the tri-bimaximal limit $T_\text{TBM}$.

In this example, which yields the Normal Ordering (NO) scenario for the light neutrino masses, the Dirac neutrino masses, $m_{D_i}$ in $D_D$, and the Majorana neutrino masses $M_i$ in $D_R$ have the following values
\begin{subequations}
\allowdisplaybreaks
\begin{align}
    D_D =& \mathrm{diag}\left(0.099, \ 3.19, \ 172.76\right) \text{GeV},
    \\
    D_R =& \mathrm{diag}\left(1.41, \ 39.2, \ 20882 \right) \times 10^{10} \, \text{GeV},
\end{align}
\end{subequations}
which, together with
\begin{subequations}
\allowdisplaybreaks
\begin{align}
|V| =&\left(
\begin{array}{ccc}
 0.974552 & 0.224127 & 0.00394435 \\
 0.224018 & 0.973733 & 0.0407452 \\
 0.00801631 & 0.0401431 & 0.999162 \\
\end{array}
\right),
    \\
    I_{CP} =& | \mathrm{Im}(V_{12}V_{23}V_{22}^*V_{13}^*)|=3.071 \times 10^{-5},
\end{align}
\end{subequations}
result in $T$ being exactly the tri-bimaximal mixing, $T= T_\text{TBM}$. We obtain also the following first light-neutrino mass $m_{\nu_1}$,  mass-differences and a PMNS matrix:
\begin{subequations}
\allowdisplaybreaks
\begin{align}
    m_{\nu_1} = 0.005125\ \text{eV} , \ \ & \Delta_{21}^2= 7.39 \times 10^{-5} \ \text{eV}^2, \ \ \Delta_{31}^2= 2.5 \times 10^{-3}\ \text{eV}^2,
    \\
    |V_{\text{PMNS}}| = &\left(
\begin{array}{ccc}
 0.811536 & 0.563176 & 0.155696 \\
 0.420431 & 0.607682 & 0.673766 \\
 0.405766 & 0.55996 & 0.722356 \\
\end{array}
\right),
\\
     \sin^2(\theta_{sol})= 0.3250,\ \ & \sin^2(\theta_{atm})=  0.4652 ,\  \ \left| V^{\text{PMNS}}_{13} \right|^2 = 0.02424.
\end{align}
\end{subequations}

\subsection*{Analysis of parameter-space}
Next, we make a vaster exploration of the region which includes all examples in agreement with the CKM and neutrino data, similar to the one just described, and the parameter-region near to these where as explained in the previous section and after Eq.(\ref{eq:leptonmass}), we have $V^\prime , W\approx V_\text{CKM}$. Here, we allow for the elements of these to differ at most $10\%$ from the elements of the CKM matrix. 

Following \cite{ParticleDataGroup:2022pth}, we assume that the sum of the light neutrino masses does not exceed the current constraint $\sum_i m_{\nu_i}<0.12 \ \text{eV}$. We fix the third Dirac mass-parameter to be $m_{D_3}=m_t=172.76\ \text{GeV}$. This is a typical choice, but it is clear from Eqs.(\ref{eq:eff}, \ref{eq:eff2}) that the subsequent results have be to appropriately re-scaled if we were to assume a different value. E.g. if we chose $m_{D_3}$ to be $10$ times smaller $m_{D_3}=17.276\ \text{GeV}$, then the corresponding values for $m_{D_1}, m_{D_2}$ would also have be $10$ times smaller while the corresponding $m_{R_i}$ would have to be $100$ times larger, in order to have the same light neutrino mass values.

Under these conditions, and scanning through parameter-space, the solutions that we have found, result in the following restrictions for the mass range for the other two neutrino Dirac mass parameters:
\begin{equation}
\begin{split}
   0.1 \  \text{GeV}\lesssim & \ m_{D_1} \lesssim 4\ \text{GeV},\\
    2  \ \text{GeV}\lesssim & \ m_{D_2} \lesssim 6 \ \text{GeV} ,
\end{split}
\label{diracrange}
\end{equation}
In addition, the Majorana neutrinos masses are subject to a limited range, where
\begin{equation}
\begin{split}
   2\times 10^{10} \ \text{GeV}\lesssim & \ M_{1} \lesssim 9\times 10^{11}\ \text{GeV}, 
   \\
   3\times 10^{11}  \ \text{GeV}\lesssim & \ M_{2} \lesssim 2\times 10^{12}\ \text{GeV},
     \\
   2\times 10^{14} \ \text{GeV}\lesssim & \ M_{3} \lesssim 1.5 \times 10^{15}\ \text{GeV} ,
\end{split}
\label{majoranarange}
\end{equation}
Very roughly our results respect the (obvious) relation, where
\begin{equation}
 \ \frac{m^2_{D_{i}}}{M_{i}} = O(m_{\nu_{i}})
\end{equation}
and we find a minimum value for the smallest light neutrino mass\footnote{In the context of NO this implies $\sum_i m_{\nu_i} \gtrsim 0.065$ \text{eV}, which is in accordance with a current experimental lower bound $\sum_i m_{\nu_i}\geq 0.06$ \text{eV} proposed for NO \cite{ParticleDataGroup:2022pth}.}
\begin{equation}
m_{\nu_{1}}\simeq 5\times 10^{-3}\ \text{eV}.
\end{equation}

\begin{figure*}
\centering
\resizebox{0.85\textwidth}{!}{%
  \includegraphics[scale=0.6]{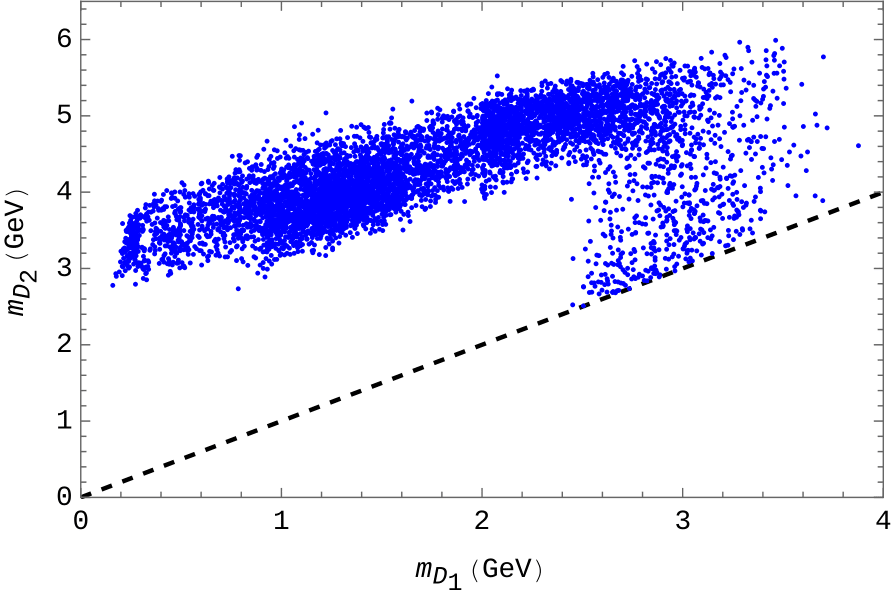}
}
\caption{We plot first two neutrino Dirac masses $M_{D_1}$, $M_{D_2}$ assuming that $m_{D_3}=m_t$ and in the ($10\%$) region where $V',W\approx V_\text{CKM}$. For simplicity, only cases where $m_{D_1}\leq m_{D_2}$ were considered.}	\label{figure1}       
\end{figure*}

\begin{figure*}
\centering
\resizebox{0.85\textwidth}{!}{%
  \includegraphics[scale=0.6]{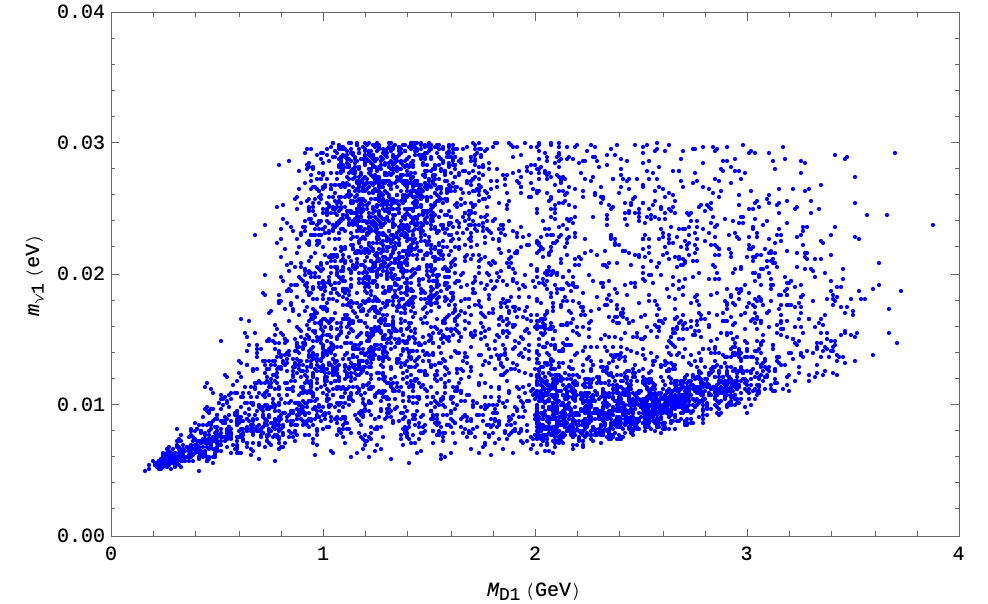}
}
\caption{We show the range of the first light neutrino mass $m_{\nu_1}$ e.g. against the first Dirac mass parameter. The cut-off at $m_{\nu 1}\approx 0.03$ \text{eV} arises from the implemented bound $\sum_i m_{\nu_i}<0.12$ \text{eV}.}	\label{firstneutrino}       
\end{figure*}

\begin{figure*}
\centering
\resizebox{1.\textwidth}{!}{%
  \includegraphics[scale=1]{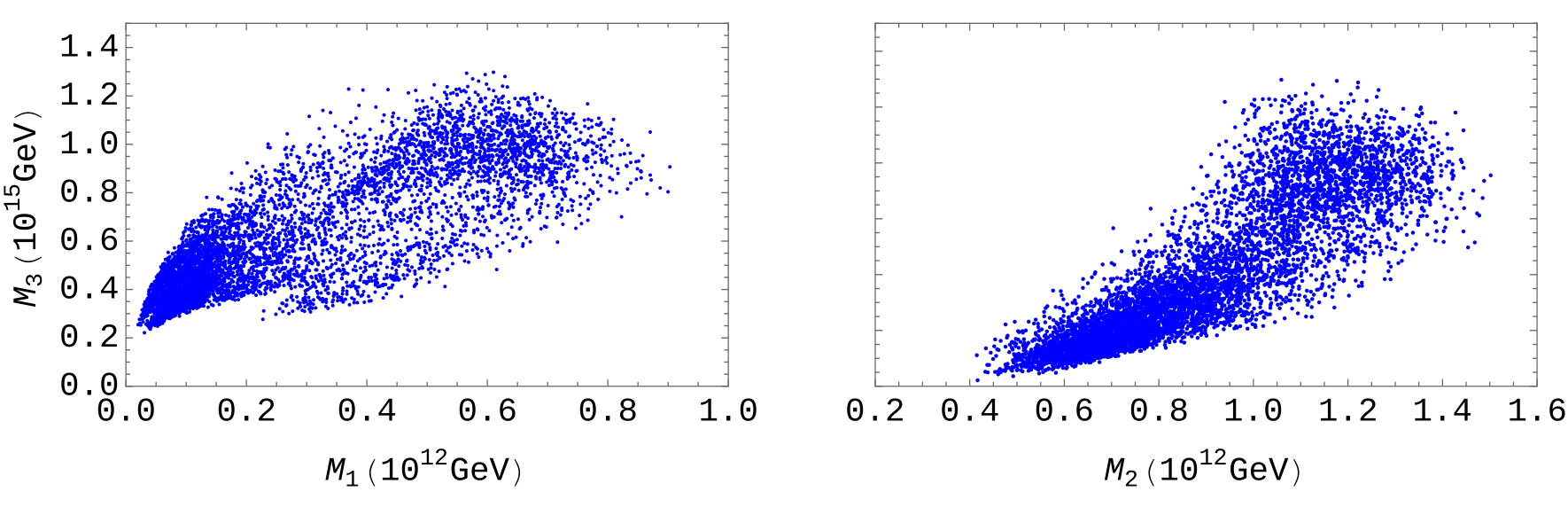}
}
\caption{Masses of the heavy Majorana neutrinos for the same conditions considered in Fig.\ref{figure1}.}	\label{figure2}       
\end{figure*}

We have also resumed our results in Figs. (\ref{figure1}- \ref{figure2}), where, more precisely, different relations are evident between diverse possible parameter combinations. 

The Majorana neutrino mass parameter spectrum, which we obtain here, is in clear contrast with the results in \cite{Branco:2020yvs}, where it is possible to obtain "heavy" Majorana neutrinos with KeV masses or even a few eV, but then for these latter masses, the Dirac mass matrix $M_D$ is totally unrelated to $V_\text{CKM}$.

It is worthwhile to mention that we have not found any solution for the case of Inverse Ordering (IO) of the light neutrino mass spectrum. Apparently, the quark-lepton CKM-PMNS relationship here described seems to prefer NO.

A final remark concerning repercussions for Leptogenisis with respect to our ansatz-proposal where 
$M_D = V D_D V^\dagger$ as in Eq. (\ref{eq:leptonmass2}). The matrix combination $M_D^\dagger M_D$ responsible for obtaining sufficient CP-violation, in the case of Leptogenisis, is in our case, also driven by the same CKM matrix $V$, where clearly $M_D^\dagger M_D=V D^2_D V^\dagger$. For our ansatz and following \cite{Branco:2002kt}, we have checked that indeed, numerically, sufficient CP-violation is possible. 

\section{Conclusions} \label{sec:conclusion}
	
In the framework of type one seesaw and assuming the principle of a lepton-quark symmetry, we have shown a specific relationship between the quark and the lepton mixing. In a weak basis where the charged lepton mass matrix and the right handed Majorana neutrino mass matrix are both diagonal and real, and assuming that the neutrino Dirac mass matrix $M_D$ is related to $V_{\text{CKM}}$, we are able to obtain the empirical King-Mohapatra-Smirnov relation, which suggests a connection between the CKM and PMNS mixing where $|V^\text{PMNS}_{13}|\approx \frac{1}{\sqrt{2}}\sin(\theta_C)$.
This is only possible within a restricted region of the Dirac and Majorana mass parameters.

We have also done an extensive numerical analysis of the available parameter space, and found that, e.g. if the Dirac neutrino masses are of the order of the EW scale or somewhat smaller, our framework makes specific predictions for the heavy Majorana neutrino masses, which must then be of the order of $10^{10}\ \text{GeV}$ or larger.
Furthermore, our scenario seems to favour the Normal Ordering.

\section*{Acknowledgments}
This work was partially supported by Fundação para a Ciência e a Tecnologia (FCT, Portugal) through the projects CFTP-FCT Unit 777 (UIDB/00777/2020 and UIDP/00777/2020), PTDC/FIS-PAR/29436/2017, CERN/FIS-PAR/0008/2019 and CERN/FIS-PAR/0002/2021, which are partially funded through POCTI (FEDER), COMPETE, QREN and EU.
	
	\providecommand{\noopsort}[1]{}\providecommand{\singleletter}[1]{#1}%
	
	\providecommand{\href}[2]{#2}\begingroup\raggedright\endgroup
		
	\end{document}